Incoherent Tunneling Amplitude in High Tc Cuprates

Philip W Anderson and Philip A Casey

ABSTRACT

**A 2004 paper of Anderson and Ong calculated the asymmetric coherent spectrum of point-contact tunneling into a Gutzwiller-projected superconductor. We here correct some details of that paper and present recipes for the many-quasiparticle, incoherent spectrum, the major portion of which results from the spin-flip decay of the injected quasiparticle. An illustrative example is shown.**

INTRODUCTION

In 2004 Ong and the present author[1] published a calculation of the tunneling density of states of a high Tc superconductor using the "plain vanilla" theory in which the ground state is modeled by a Gutzwiller projected BCS wave function. We showed that the overall spectrum is markedly asymmetric between holes and electrons but that the "coherence peaks" observed in tunnelling, which reflect states actually at the gap energy, are symmetric. In that paper we neglected the part of the tunneling amplitude which we called "incoherent" because it comes from three-Fermion terms in the commutator of the projector with the Fermion operator. The basis for doing so was correctly stated in principle for hole tunneling, that the amplitude might be small because it is mostly above $3\Delta$ and is distributed over the whole band. But for electrons the completely incorrect statement was made —see p ( 3)— that this amplitude is exactly zero. As we shall shortly show, in fact the relative amplitude of "coherent" and "incoherent" parts is practically the same for both signs of carriers. Davis[2], accepting our incorrect statement about the electron side, has done an elaborate analysis of the incoherent amplitude on this side assuming that it comes from modulation of the gap by a bosonic

"glue"; the message of the present paper is that we apologise for the error in our previous theory and that that analysis does not afford evidence for a glue which modulates the energy gap. It is intrinsic to the Green's function of a Gutzwiller-projected system. Yazdani[3] and others[4] have also observed the "hump" in electron amplitude; it was Yazdani's insistence that it can be symmetrized with the hole side by taking ratios to the normal state that stimulated some of the present work. Our present view agrees with this recipe.

We are by no means the first to attribute the hump structure seen in tunneling to coupling to magnetic modes: Zasadinski[5] and (discussing ARPES data, which is less straightforward) M Norman[6] have suggested that structures ascribed by others to phonons or other "glues" are compatible with coupling to the neutron collective mode. What is new here is that we are proposing a specific, novel, mechanism for this coupling and making a definite quantitative prediction.

One remark is worth making about the Davis-Balatsky paper. These authors observe an isotope effect on the frequency and therefore identify the relevant bosonic mode as a phonon. Recent, and presumably reliable, measurements have found that the isotope effect on the superfluid density, which is simply proportional to t and has no relation to the interaction, is similar to that on Tc. Clearly, then, since J is roughly proportional to $t^2$, there should be an isotope shift of the neutron resonance of similar order. We suggest that isotope measurements have no relevance, in the cuprates, to mechanistic questions, and can, and have, only confused the picture.

The other misjudgment in ref [1] is the presumption of smallness and of relevance only at high frequency of these incoherent terms. Already in that paper we remarked that they had the potentiality of bringing in some amplitude from the prominent "neutron resonance" feature in the magnetic susceptibility (on the hole side only, we assumed incorrectly). It appears that the interactions which cause the resonance lower the frequency of this term to well below $3\Delta_{max}$ and enhance its magnitude, so that the spin-flip term can dominate the incoherent spectrum.

In recent papers[7] the present author has presented a simplified version of the formalism of ref [1] which is called the 'hidden Fermi liquid" technique. This technique gives a reasonably good account of the "strange metal" which appears in the normal state where the BCS gap has been rendered ineffectual by thermal fluctuations. In the strange metal the incoherent amplitudes which we discuss here dominate the spectrum completely because of the "infrared catastrophe" at the Fermi surface, and there is no coherent quasiparticle. When a superconducting gap opens up there is no longer a Fermi surface nor an infrared catastrophe, as was shown long ago by Y Ma[8], and the incoherent amplitudes appear as 3- or more-Fermion decay amplitudes. These amplitudes peak near the gap+resonance energy and provide some of the "hump", but on the hole side there is also an "Anderson-Ong" contribution to the apparent hump in tunneling conductivity.
.

FORMALISM

As in reference 1, we will confine our efforts mostly to calculating the STM (presumed single-site) tunneling spectrum. In reference 4, we showed that the momentum-specific and site-specific spectra are closely related, and contrary to remarks in ref [1], they have the same coherence properties because of Huygens' principle.

The central idea of the HFL method is that the Plain Vanilla[9] gap equations are the Hartree-Fock-Bogolyubov variational equations which determine the energies and wave-function parameters which apply to an *unprojected* product wave function

$$\Phi = \prod_k (gu_k + v_k c^*_{k,+} c^*_{-k,-}) |vacuum\rangle, \quad [1]$$

while the true wave function is the projected function

$$\Psi = P\Phi, \text{ where} \quad [2]$$

$$P = \prod_{sites\ i} (1 - n_i^\uparrow n_i^\downarrow)$$

From c and c* we can construct the low single-particle excitations of the "unprojected" system:

$$\gamma_{k\sigma} = u_k c_{k\sigma} + v_k c^*_{-k,-\sigma}, \quad [3]$$

and [1] can be seen to be a BCS product function (modified by an overall fugacity factor) using these excitations. The Hamiltonian which they satisfy is the *projected* t-J Hamiltonian, whose effective kinetic energy is reduced as described in ref 1 for the effects of the elimination of double occupancy.

The space defined by arbitrary occupancy of these excitations is necessary very overcomplete for the description of a general state in the lower Hubbard band; but in deriving the gap equations we are assuming they are occupied only a few at a time, and for such states one may show using momentum conservation that they are not overcomplete. In the normal state, they form a pair of independent Fermi liquids, one for each spin, but in the superconducting state, the observation of the magnetic resonance shows that they can interact appreciably. We will call them pseudoparticles of the hidden Fermi liquid.

These excitations are not quasiparticles, in that they are not adiabatic continuations of the free particles of the noninteracting case. When we inject or remove a particle as in ARPES or tunneling, or when a particle is accelerated by an external field, it acts *outside of the projector* and the Fermion operator must be commuted through the projector in order to express it in terms of its action on the pseudoparticles which are the actual eigenexcitations of the hidden Fermi liquid.

The most straightforward way to express this relationship is to define "hat" operators for the physical particles, projected operators which act only in the lower Hubbard band, but represent the effect of adding real physical particles or holes in that band.

$$\hat{c}^*_{i,\sigma} = P c^*_{i,\sigma}\ P = (1 - n_{i,-\sigma}) c^*_{i,\sigma} = c_{i,-\sigma} c^*_{i,-\sigma} c^*_{i,\sigma} \quad [4a]$$

note however that this operator is not the full normalized electron creation operator; that operator has matrix elements into the forbidden subspace of the upper Hubbard band. Thus the tunneling DOS due to $\hat{c}^*$ is $2x/(1+x)$ (approximately) of that due to $\hat{c}$. We may write the latter as

$$\hat{c}_{i,\sigma} = c_{i,\sigma} c_{i,-\sigma} c^*_{i,-\sigma} \quad [4b]$$

When we examine the "hat" operators of equation [4], we note that they can be factorized in two different manners:

$$\hat{c}_{i\sigma} = c_{i\sigma} c_{i-\sigma} c^*_{i-\sigma} = c_{i-\sigma} S^-_i \text{ or } = c_{i\sigma}(1 - n_{i-\sigma}) \quad [5]$$

When we write these two factorizations out in momentum space they look rather different.

$$\hat{c}_{k,\sigma} = N^{-1/2} \sum_j e^{ir_j \cdot k} \hat{c}_{j,\sigma} = N^{-1} \sum_q c_{k-q,-\sigma} \sum_{k'} c^*_{k'-q,-\sigma} c_{k',\sigma}$$

$$= \sum_q c_{k-q,-\sigma} S_q^-$$

$$or = N^{-1} \sum_q c_{k-q,\sigma} \sum_{k'} c_{k',-\sigma} c^*_{k'-q,-\sigma} = \sum_q c_{k-q,\sigma} (-\rho)_q \qquad [6]$$

These two factorizations are appropriate in different regions of the summed momentum and spin variables. The basic idea of the Luther-Haldane tomographic desription of the kinematics of the Fermi liquid comes from two observations about the effect of the exclusion principle and conservation of energy on electron-electron scattering for states close to a Fermi surface. One is that as we squeeze down to the Fermi surface only nearly forward scattering survives (see fig 6.1 of my book[10].) It is this forward scattering which causes, for instance, the mean-field interactions in Fermi liquid theory. The second observation is that inelastic scattering only conserves energy when the Fermi velocities match(since the energy difference is $(v_F - v_F') \cdot q$ ) . Thus for a Fermi liquid-like system inelastic scatterings of particle near the Fermi surface can be thought of as generating tomographic Tomonaga bosons, with the incoming particle losing momentum q along the direction of its velocity and generating a Tomonaga boson composed of soft electron-hole pairs of that momentum. The two spin channels in which this can occur are the two versions of equation [6]. Both give a singular contribution to the renormalization constant and their sum causes the "strange metal" phenomena observed above Tc, as described in reference [5].

When the gap opens up below T* (below Tc for optimal doping) there is no longer a Fermi surface and the singular power laws of the "strange metal" disappear as described in reference [6], and are replaced by coherence peaks plus backgrounds given by perturbative power series which have amplitudes above 3Δ for the first term, 5Δ for the second, etc. We have calculated these and (as was actually foreshadowed in ref [1]) they are not appreciable; they certainly do not account for the backgrounds seen in Dessau's ARPES experiments or the "humps" seen in the tunneling data.

What we are forced to conclude is that the assumption of free, noninteracting pseudoparticles which is successful in describing the strange metal phase is not satisfactory for the superconducting phase. Fortunately, one aspect of the superconducting phase is already known to imply strong interactions between the quasiparticles, namely the magnetic susceptibility as measured by neutron scattering. Neutron scattering exhibits a pronounced magnetic resonance in the superconducting state in the neighborhood of $qa=\pi,\pi$, which lies in an energy-momentum range quite outside of and below the particle-hole continuum; and the magnetic spectrum of the superconductor, while otherwise lying more or less within the continuum boundaries, seems to be considerably enhanced at low energies and distorted by interactions.

The interaction which is responsible here is of course the superexchange interaction J of the t-J Hamiltonian, which causes both superconductivity and the antiferromagnetism of the insulator. J is actually somewhat enhanced in the Gutzwiller-projected state relative to the free Fermi liquid. The magnetic resonance noted above is well understood to be the "soft mode" of the transition from the d-wave state to incipient antiferromagnetism. This mode is low and strong in the d-wave state because the superconducting coherence factors are favorable for $q=\pi,\pi$; one can expect the spectrum in the normal state to be much less distorted and to approximate the free quasiparticle Pauli susceptibility.

To get a rough approximation of the effect of magnetism on the single-particle spectrum it may suffice to use the factorizations given in equations [5] (for tunneling) and [6] (for ARPES), together with empirical data on the susceptibility from neutron measurements. We shall carry out the former here and then make a few remarks about the ARPES data.

Our approximation, as yet not rigorously justified, is to assume that the incoming quasiparticle creates a spin or charge density wave plus an ongoing pseudoparticle, and that these entities do not further interact. It is reasonable in that their propagation velocities do not match at all; there is no obvious reason why spin and particle number should not propagate roughly independently.

If this is the case, we can factorize the Greens function as we did in the strange metal calculations:

$$G(0,t) = \langle 0 | \hat{c}_{i,\uparrow}{}^*(t) c_{i,\uparrow}(0) | 0 \rangle$$
$$= G_{coherent} + G_{inc,density} + G_0(0,t) \langle 0 | S_i^+(t) S_i^-(0) | 0 \rangle \quad [7]$$

Here $G_{coherent}$ is the Green's function as calculated in reference [1], using the average value of $(1-n_{-\sigma})$ in the definitions of the hat operators. $G_{inc, density}$ is the rather small contribution to background from the Ma power series, which begins at $3\Delta$. $G_0$ is essentially the same as $G_{coherent}$ except that it need not be corrected for $(1-n_\sigma)$; it is the propagator of the down-spin pseudoelectron. Its Fourier transform will contain the tunneling "coherence peak" at $\omega = \Delta_{max}$.

The spin-spin correlation function in [6] is of course the Fourier transform of the spin susceptibility:

$$\int dt e^{i\omega t} \langle 0 | S_i^+(t) S_i^-(0) | 0 \rangle = \chi_i''(\omega) = N^{-1} \sum_k \chi''(k,\omega) \quad [8]$$

Keimer et al[11] have compiled precisely the sum in the last term in [7] for the cuprate YBCO6.6, unfortunately not totally comparable to the somewhat more highly doped BISCO samples on which there is good tunneling data. The energy-dependence is given in a figure from their paper which we reproduce as our figure [1]:

**Figure 1: Local Susceptibility as a Function of Energy ForYBCO$_{6.6}$ (from ref[9].)**

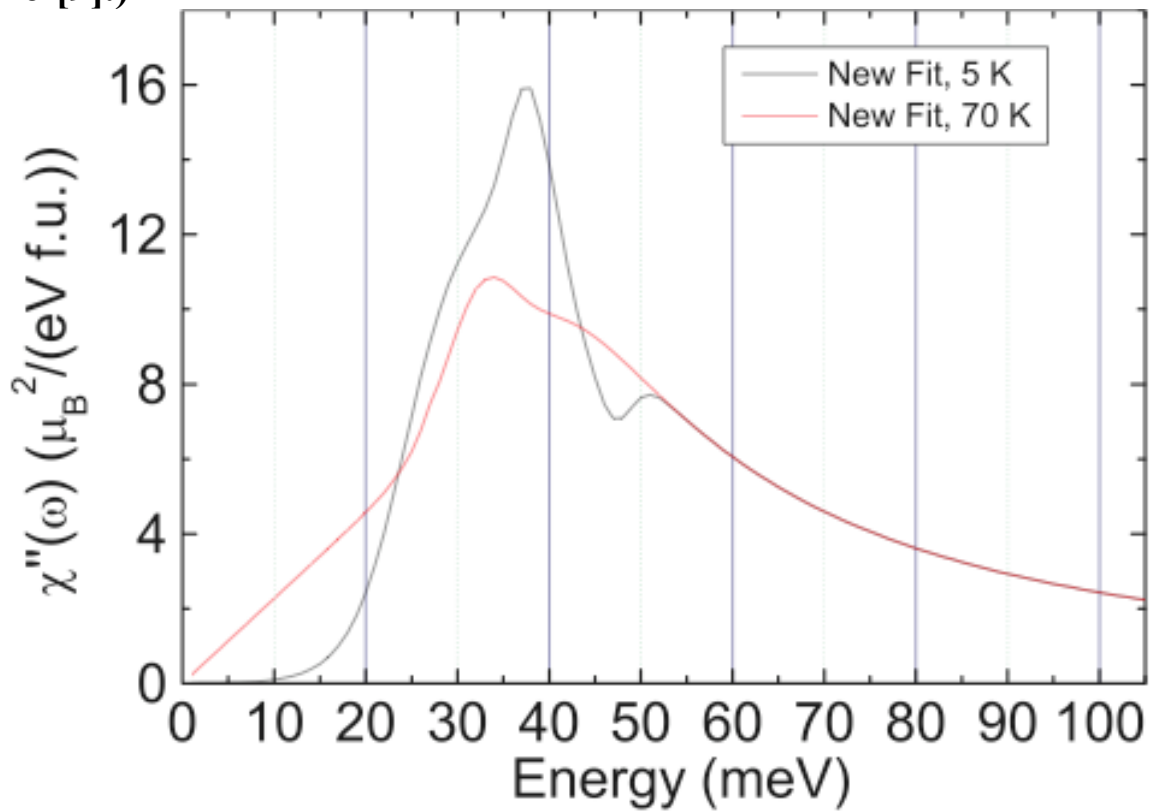

The resonance does not seem to vary qualitatively from one cuprate to another except for LASCO, which for some reason has never exhibited it at least in fully developed form. The remarkable thing which can be calculated from figure 1 is that indeed the spectral shift downwards of the local susceptibility is large. The integral of fig 1 is approximately .5 $\mu_B^2$, in other words half of the susceptibility shows up below 100mev, where it should in a free Fermi gas be evenly distributed over the bandwidth; and in an s-wave state it would be suppressed in the same region. This must mean that, as argued by Scalapino[12], some appreciable fraction of the pair binding energy is caused by the softening of spin fluctuations in the d-wave state. How much can be estimated from the data of fig 1 but we don't do that here.

We can calculate the fourier spectrum of the last term in [6] by convoluting $\chi''$ with $G_0$:

$$\mathrm{Im} G(\omega) = \mathrm{Im} G_{coherent} + \mathrm{Im} G_{inc\ density} + \int d\omega' G_0(\omega - \omega') \chi''(\omega') \quad [8]$$

It is only a rough estimate to convolute fig 1 with the coherence peak from a near-optimal BISCO sample[13], but it can tell us that the background "boson" peak is indeed the right sort of shape and magnitude to fit the last term of equation [8]. The result of such a procedure is shown in fig 2 and compared with the actual spectrum.

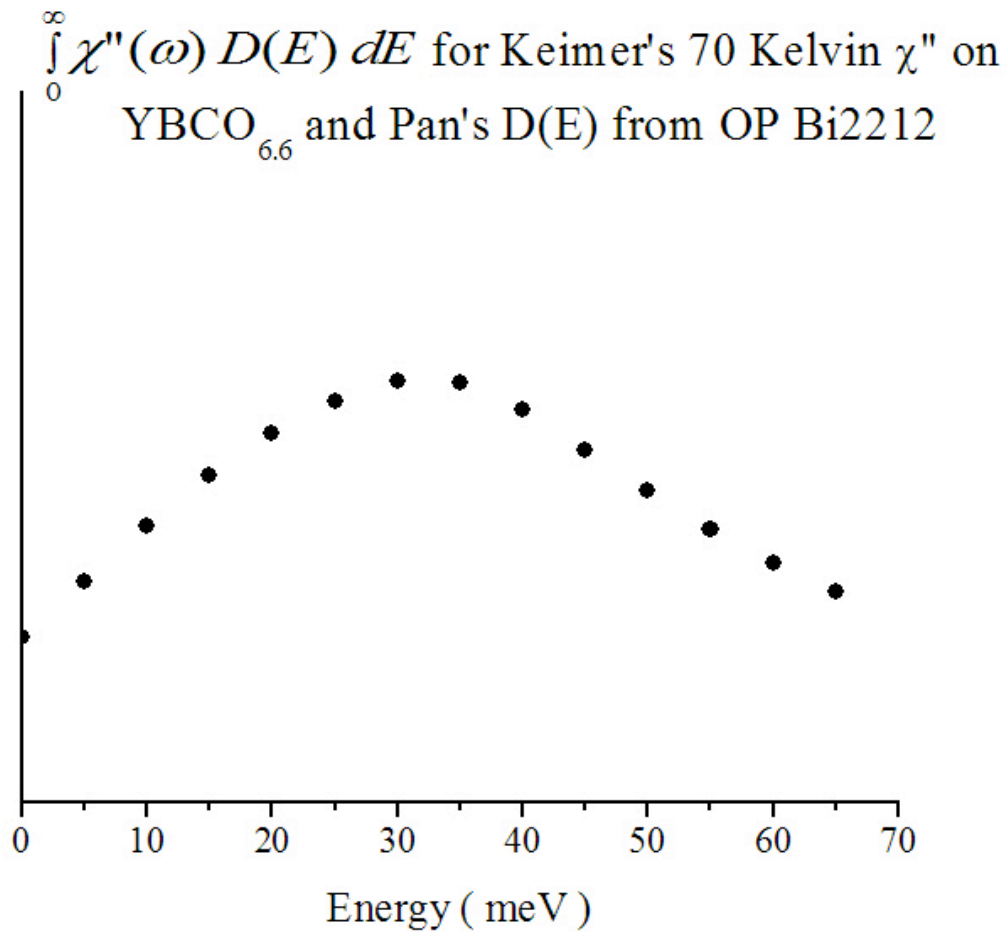

**Figure 2.** "Hump" in tunneling spectrum estimated from Hinkov data. The energy is measured from the coherence peak which should be imagined to be added in around E=0 with a total area comparable with that of the hump.

Unfortunately, to estimate the spin-fluctuation background for a particular ARPES spectrum requires more detailed information about the q-dependence of the susceptibility and also equally detailed consideration of the pseudoparticle dispersion. One fact which is immediately clear is that there is no expectation that the background should be universal. Specifically, near the nodal point there are no low-energy states displaced by π,π in momentum so the resonant hump will not appear, as is indeed observed by Dessau et al,[14] but as we move along the Fermi surface toward the zone corner the resonance should come into play.

CONCLUSIONS

We have found that the Hidden Fermi Liquid idea gives a good account of ARPES and tunneling spectra in the normal (strange metal) state of the cuprate superconductor, but in the superconducting state laser ARPES and tunneling data, especially the former, show an appreciable background with a prominent "boson" hump. We have shown that it is at least reasonable to believe that these backgrounds are caused by the same quasiparticle decay process as in the HFL theory of the normal state, but now the decay occurs into spin fluctuations which are enhanced by the proximity of antiferromagnetism. This mechanism for the structure must not be confused with the Eliashberg phenomenon observed in conventional superconductors, which occurs only because of modulation of the energy gap; no such modulation is assumed here and no implication as to the source of the superconducting gap may be drawn.

ACKNOWLEDGEMENTS
We are glad to acknowledge the sharing of their data by B Keimer and V Hinkov, and extensive discussions with A Yazdani. We have also benefited from discussions with J C Davis, M Norman and D Dessau.